\documentstyle[prl,aps,twocolumn,epsf]{revtex}

\title{Quantum noise and correlations in
resonantly enhanced wave mixing based on atomic coherence}

\author{M.~D.~Lukin$^{1,2,3}$, A.~B.~Matsko$^{2,3}$, M.~Fleischhauer$^{2,4}$, and 
M.~O.~Scully$^{2,3}$}

\address{$^1$ ITAMP, Harvard-Smithsonian
Center for Astrophysics, Cambridge, MA 02138,}
\address{$^2$Department of Physics, Texas A\&M University,
         College Station, Texas 77843-4242,}
\address{$^3$Max-Planck-Institut f\"ur Quantenoptik, 
Garching, D-85748, Germany,}
\address{$^4$Sektion Physik,  
Universit\"at M\"unchen, 
D-80333 M\"unchen, Germany}
  
\begin{document}

\maketitle

\begin{abstract}
We investigate the quantum properties of fields generated 
by resonantly enhanced wave mixing
based on atomic coherence in Raman systems.
We show that such a
process can be used for  generation of pairs of Stokes and
anti-Stokes fields with nearly perfect quantum correlations, yielding 
almost complete (i.e. 100\%)
squeezing without the use of a cavity.
 We discuss the extension of the wave mixing
interactions into the domain of a few interacting light quanta.
\end{abstract}


\narrowtext

\

One of the intriguing and potentially useful aspects of 
nonlinear optical phenomena is their ability to suppress intrinsic
quantum fluctuations \cite{walls-book}. However, the efforts to 
exploit  these properties
were hindered, either by the small values of nonlinearities in available 
optical crystals,
or  by absorption losses 
and the associated noise 
in resonant atomic systems with large nonlinearities. 
For example, four-wave mixing is known to
result, in principle,  in squeezed-state generation or non-classical 
correlations \cite{4WM_squeeze}, but all experimental realizations 
reported to date showed rather limited noise reduction 
and required the use of cavities \cite{4WM_squeeze_exp}.

The work of the past few years has shown that 
substantial improvements in resonant nonlinear optics can be achieved by 
utilizing the concepts of quantum coherence and
interference \cite{harris97phys.today,hemmer95}. 
The aim of the present contribution is to demonstrate the 
usefulness of this regime 
of nonlinear optical enhancement for applications involving quantum 
correlations and reduction of quantum noise.   
As an example, we consider here four-wave mixing in resonant Raman
systems \cite{hemmer95}, where atomic phase coherence can be used to generate
a large nonlinearity and at the same time suppress resonant absorption.
Recent theoretical \cite{lukin98prl} and experimental work 
\cite{zibrov98}  demonstrated that the efficient nonlinear 
interactions in this system can lead to 
mirrorless parametric oscillation, where pairs of counter-propagating 
Stokes and anti-Stokes
photons are generated spontaneously from noise.  
We here show that under certain, very realistic conditions this 
 process
can  be considered as ideal from the viewpoint of quenching of 
quantum noise.
As a result, the generated Stokes and anti-Stokes field
components can possess practically perfect quantum correlations, 
 leading e.g. to an almost complete suppression of the quantum 
fluctuations in one quadrature of a
combined mode (i.e.  100\% squeezing).  
We point out  that this  
can be achieved even in the case when the intensity of the driving
 fields approaches, under realistic experimental conditions, the
 few-photon level. These results, together with recent studies on strongly 
interacting photons  \cite{imamoglu97}, single-photon switching \cite{harris98}
and few photon quantum control
\cite{quantum_control}, show that a truly new regime of nonlinear optics
involving just a few interacting light quanta is feasible.

Physically, such a  performance of the nonlinear
media is due to the possibility of eliminating the resonant absorption and 
associated noise processes via atomic coherence. 
 Furthermore, the associated large linear dispersion 
is very important for achieving phase matching
\cite{zibrov98} and plays a key role in the reduction of the oscillator
linewidth \cite{lukin98ol}
which will be discussed in detail elsewhere. 
The present results open new interesting possibilities for
applications as diverse as novel frequency standards and gravity-wave
detection  
on one hand and quantum-information processing
 on the other \cite{kimble}.

In the present paper we discuss the noise properties of 
electromagnetic waves propagating in a medium consisting of double-$\Lambda$ 
atoms  as shown in Fig.1 in a four-wave mixing configuration. 
Four optical waves 
are tuned to the vicinity of
the corresponding optically  allowed transitions. 
These fields include two counter-propagating 
driving fields with frequencies 
$\nu_{d1}$, $\nu_{d2}$ and Rabi-frequencies $\Omega_{1}$ and $\Omega_{2}$, 
and two probe fields (anti-Stokes and Stokes)  with carrier frequencies 
$\nu_1 = \nu_{d1}+
\omega_0$ and $\nu_2=\nu_{d1}-\omega_0$, where $\omega_0=\omega_{b1}-
\omega_{b2}$ is the ground-state frequency splitting. 
The probe fields are described quantum mechanically
in an effective 1-D model. 
  The
fields interact via the long-lived coherence on the
dipole-forbidden transition between the metastable ground states 
$|b_1\rangle$ and $|b_2\rangle$.

We here utilize a  Langevin approach  
in which collective atomic variables and fields
are described by  
time- and position-
dependent stochastic differential equations
with $\delta-$ correlated 
Langevin forces\cite{fleischhauer95pra}. 
The present approach develops from   
the semiclassical analysis of Ref.[6], 
in which, in particular,  a phase transition to mirrorless 
parametric oscillation was noted. We now proceed with an analysis
of the quantum fluctuations in such a system. 
We obtain  stochastic equations which, for undepleted pump fields,
can easily be solved by Fourier-transformation.

We introduce slowly-varying dimensionless field
variables ${\hat E}_{1,2}(z,t)$ which  contain only modes propagating
in the $+z$ and $-z$ direction respectively 
\begin{eqnarray}
{\hat E}_1(z,t) &=& \sum_{k>0} a_k(t)\, {\rm e}^{i(k-k_1)z}
\, {\rm e}^{i\nu_1 t},\\
{\hat E}_2(z,t) &=& \sum_{k>0} a_{-k}(t)\, {\rm e}^{-i(k-k_2)z}
\, {\rm e}^{i\nu_2 t},
\end{eqnarray}
where $k_{1,2}=\nu_{1,2}/c$.
Following the approach of \cite{fleischhauer95pra} we 
derive stochastic differential equations for c-number
analogues of the fields $\hat E\to E$ and 
collective atomic operators
of the medium consisting of the four-state atoms shown in Fig.1. 
Apart from the 
stochastic noise sources these equations have a form identical to the 
semiclassical 
density matrix equations for such an atomic system. The 
diffusion coefficients 
for noise correlations are derived using the 
fluctuation-dissipation theorem and
generalized Einstein relations. We find that the 
propagation of the Fourier-components of
Stokes and anti-Stokes fields 
$E_1^*(z,\omega)$, $E_2(z,\omega)$ is governed by the
(c-number) equations
\begin{eqnarray}
\left(\frac{\partial}{\partial z} +\frac{i \omega}{c}
\right) E_1^*(z,\omega) &=& -i\frac{k_1}{2\epsilon_0} P_1^*(z,\omega), \\
\left(-\frac{\partial}{\partial z} +\frac{i \omega}{c}
\right) E_2(z,\omega) &=& i\frac{k_2}{2\epsilon_0} P_2(z,\omega).
\end{eqnarray}
Here the $P_l$'s are the c-number variables proportional to the 
corresponding polarizations in the appropriate units,
and the Fourier-transform is defined as
$F^{(*)}(\omega)=1/\sqrt{2\pi}\int dt\, {\rm e}^{-i\omega t}F^{(*)}(t)$.  
Solving the equations of motion for the 
atomic variables in lowest order of the Stokes and anti-Stokes
fields we find: 
\begin{eqnarray}
&&P_{l}(z,\omega)= 
\epsilon_0\,\chi_{ll}(z,\omega) E_l(z,\omega)\label{polar1}\\
&&\quad + \epsilon_0\,\chi_{lm}(z,\omega) 
\, e^{i\Delta \vec{k}\cdot \vec{r}}\,  
E_m^*(z,\omega) - f_l(z,\omega)/k_l, \nonumber
\end{eqnarray}
where $\{ l,m \} = \{ 1,2 \}$ and $m\ne l$. $\Delta \vec{k}$ is
a possible geometrical phase mismatch, $\chi_{ll}(z,\omega)$ 
are the self-coupling and 
$\chi_{lm}(z,\omega)$ ($m\ne l$) the cross-coupling
($\chi^{(3)}$-type) susceptibilities of the medium \cite{lukin98prl}.
Both do not depend on the amplitudes of Stokes and anti-Stokes fields
but are functions of the drive-fields and thus in general space-dependent.
$f_{1,2}(z,\omega)$ are noise 
sources, which are $\delta$-correlated in frequency and
position. 
In the following we assume $\Delta \vec k=0$. Note, however, that
a non-zero phase mismatch can easily be compensated in the present system
by a small detuning of the Stokes and anti-Stokes fields from two-photon
resonance. Thus the equations of motion 
for the Fourier-components of the fields at frequency $\omega$ 
are: 
\begin{equation} \label{max} 
\frac{d}{d z}\left[\matrix{ E_1^* \cr E_2}\right]
=i\left[\matrix{a_{11} & a_{12}\cr a_{21} & a_{22}}\right]\left[\matrix
{E_1^*\cr E_2}\right]+\frac{i}{2\epsilon_0}\left[\matrix{f_1^*\cr f_2}\right],
\end{equation}
where $a_{1j}\equiv
a_{1j}(z,\omega)=-k_1 \chi_{1j}^*(z,\omega)/2 - \delta_{j1}\,\omega/c
$, and $a_{2j}\equiv
a_{2j}(z,\omega)=-k_2
\chi_{2j}(z,\omega)/2 +  \delta_{j2}\,\omega/c\,$.  

In order to solve the inhomogeneous boundary-value problem 
we assume undepleted driving fields 
and transform away 
their remaining space dependence due to the 
 refractive index. Thus $a_{ij}(z,\omega)\to a_{ij}(\omega)$.
Assuming 
vacuum input ($E_1^*(0) = 0, E_2(L) =
0$) at both sides of the medium of length $L$
we eventually find: 
\begin{eqnarray} 
\label{sol1} 
E_1^*(L) &=&  
\int_0^L \!\! dz'\,  
\, \frac{ i f_1^*(z') M(z') \eta
+ \,a_{12}\, f_2(z')\, {\rm sin}( \eta z')}
{2 \epsilon_0 M(L) \eta  e^{ i \tilde a(z'-L)}},\\ 
\label{sol2}
E_2(0) &=& 
 \int_0^L\!\! dz'\,  
\, \frac{ 
\, a_{21}\, {\tilde f}_1^*(z')\, {\rm sin}( \eta z') -
i {\tilde f}_2(z')
M(z')\eta}{ 2 \epsilon_0 M(L) \eta e^{i \tilde a (L-z')}
},
\end{eqnarray}
where we have dropped the frequency dependence.
${\tilde f}_i(z) = f_i(L-z)$, $\tilde a = (a_{11}+a_{22})/2$, 
$ a = (a_{22}-a_{11})/2$,  
$\eta = \sqrt{a^2 + a_{12}a_{21}}$, and $M(z,\omega) = {\rm cos}
( \eta z) + i\, a/\eta\, {\rm sin}(\eta z)$. 
These expressions predict  infinite  growth of the
Stokes and anti-Stokes fields from vacuum when 
\begin{equation}
M(L,\omega)\to 0\qquad {\rm or}\qquad {\rm tan}(\eta L) = 
i\frac{\eta}{a},
\end{equation}
which is the oscillation condition \cite{lukin98prl,remark1}.

Let us proceed now with a special case in which one of the driving 
fields (say $\Omega_1$) is tuned near resonance with the corresponding 
single-photon transition $b_2 \rightarrow a_1$, whereas  the second
driving field $\Omega_2$ has a detuning $\Delta \gg |\Omega_2|$ from the 
transition $b_1 \rightarrow a_2$. For simplicity 
 assume also equal Rabi-frequencies of the driving fields 
$|\Omega_1|=|\Omega_2|=|\Omega|$. 
In this case most of the population is in the lower state $b_1$, 
and there is almost no absorption of the
driving fields.  For small 
Fourier frequencies (close to the two-photon resonance) we 
find :
\begin{eqnarray}
a_{11} = \frac{\kappa\, \bigl[-\omega+i\gamma_0\bigr]}{|\Omega|^2}- \displaystyle 
\frac{ \omega}{c}, 
a_{12} = a_{21} = \displaystyle 
\frac{{ \kappa }}{\Delta},  
a_{22} =  {\omega \over c}. 
\label{part2}
\end{eqnarray}
We here have assumed that 
the coupling constants $\kappa = 3/(8\pi)
(N\lambda_i^2 \gamma_a)$ are equal for all transitions and that 
the two-photon detuning of all relevant Fourier components 
is small, such that $| \omega / \Omega| \ll 1$. 
$\gamma_a$ is the common decay rate out of the upper levels,
$\gamma_0$ is the decay rate of the coherence 
between the lower states, and
$N$ is number density of atoms. Following the procedure of 
Ref.\cite{fleischhauer95pra},
we find for the non-vanishing noise correlations\cite{remark}
\begin{eqnarray} 
\langle
f_1(z,\omega)f^*_1(z',\omega') \rangle &\simeq& 4 \epsilon_0^2  {\kappa L 
\over c}  
{\gamma_0 \over
\Omega^2} \delta(z-z') \delta(\omega+\omega'),  \\
\langle f_1(z,\omega) f_2(z',\omega') \rangle &\simeq&  4 \epsilon_0^2  
{\kappa L \over c} { i \over \Delta} \delta(z-z') \delta(\omega+\omega'),
\\
\langle f_2(z,\omega) f^*_2(z',\omega') \rangle &\simeq&  4 \epsilon_0^2  
{\kappa L \over c}  {\gamma_a \over
\Delta^2} \delta(z-z') \delta(\omega+\omega'),
\label{noise}
\end{eqnarray}
where we have identified the quantization length with the
length of the cell $L$. 
In the case described by Eq.(\ref{part2}) parametric oscillation 
occurs at $\omega = 0$ when 
\begin{eqnarray} 
M(L,0) = {\rm cos}(\eta L) + (\gamma_{0} \Delta)/(2 |\Omega|^2) 
{\rm sin}(\eta L) =0, 
\end{eqnarray} 
with $\eta = \kappa \sqrt{1/\Delta^2- 
\gamma_0^2/4|\Omega|^4}$. Hence oscillation can be achieved
 if $\Omega^2 > \gamma_0|\Delta|/2$. It should be noted that this is
easily satisfied since $\gamma_0$ is the relaxation rate of a long-lived
ground-state coherence.
Close to this oscillation condition the spectrum of the
output field 
diverges \cite{remark1}
\begin{eqnarray} 
n_i (\omega) \equiv  \frac{c}{L}
\int d \omega' E_i^*(\omega) E_i(\omega') \sim {1 \over 
|M(L,\omega)|^2}. 
\label{numb}
\end{eqnarray}

Note that in the limit $\gamma_0\rightarrow 0$, $\Delta \gg \gamma_a$ and  
$\omega \rightarrow 0$ the coefficients  $a_{11}, a_{22}$, which 
correspond to the self-coupling
 susceptibilities, become negligible and all noise  
correlations except for $\langle f_1 f_2 \rangle$ vanish. This corresponds to 
four-wave mixing with {\it ideal
noise properties} \cite{walls-book,4WM_squeeze}. 
The Stokes and anti-Stokes photons 
generated from vacuum
($E_1^*(0) = 0, E_2(L) = 0$) possess in this case  perfect
quantum correlations, i.e. 100\% squeezing. 
Hence the quantum fluctuations of a particular
quadrature of the linear combination of output fields 
${\hat d_\theta(\omega)} \equiv ({\hat E}_1(L,\omega) + {\hat E}_2(0,\omega))
e^{i\theta}/\sqrt{2}$ 
can be almost completely suppressed near the threshold of parametric 
oscillation. 
We define the fluctuation spectrum of the combined mode 
at the output of the cell by
\begin{equation}
S_\theta(\omega)  = {c \over 4 L} \int d \omega' \bigl\langle
 [{\hat d}_\theta(\omega) + 
{\hat d}_\theta^\dagger(\omega)], [ {\hat d}_\theta(\omega') + 
{\hat d}_\theta^\dagger(\omega')]\bigr\rangle, 
\end{equation}
where $\langle a,b \rangle  = \langle ab \rangle - \langle a \rangle
\langle b \rangle$.
As can be verified from the 
commutation relation $[{\hat E}_{1,2}(z,\omega),{\hat E}_{1,2}^\dagger
(z,\omega^\prime )]= (L/c)\delta(\omega-\omega^\prime)$ (which holds for 
Fourier-frequencies small compard to the carrier frequencies),
the normalization is such that $S_\theta = 1/4$ 
corresponds to the standard quantum limit. Using
Eqs.(\ref{sol1},\ref{sol2}) for the evaluation of normally ordered 
averages, and
assuming that the system is close to the threshold 
($|M|^2 = |M(L,0)|^2 \ll 1$)
we find for the optimum phase $\theta=\pi/4$
\begin{equation}  
S_+(0)\equiv S_{\pi/4}(0)=
\frac{|M|^2}{4}+  {\pi \over 4} \left(2 
\frac{ \gamma_{0} \Delta}{ |\Omega|^2} 
+ 
\frac{ \gamma_a}{\Delta} 
\right ),
\label{spect}
\end{equation}
where we have neglected by all but linear 
terms in $\gamma_0/\Omega^2$ and $\gamma_a/\Delta^2$. 
The first term on the rhs of the above expression is the residual quantum 
noise supressed by to nonlinear wave-mixing. The second term is an
atomic noise contribution, which results from the finite relaxation rate 
of the ground state coherence and the associated absorption losses.
Finally, the third contribution is the corresponding 
 noise contribution due to the absorption of the 
far-detuned driving field. Choosing the optimum value for the detuning
($\Delta^2_{opt} = \gamma_a|\Omega|^2/(2\gamma_0)$) 
we find that the maximum noise suppression is reached already before the 
oscillation threshold (for $|M|^2 < \sqrt{\gamma_0 
\Delta/\Omega^2}$), and
is given by:
\begin{equation}\label{noise1}
S_+(0)
\rightarrow  \pi   \left( \gamma_{0}\gamma_a \over 2 |\Omega|^2 
\right)^{1/2}.
\label{ideal}
\end{equation}
The extent to which the
parametric oscillator can be considered as ideal is 
determined by the absorption losses of the medium.
In contrast to the usual two-level type systems \cite{4WM_squeeze_exp}  
this absorption
is here determined by the 
decay of the ground state coherence and by 
the detuning of one of the
driving fields from single photon resonance ($\Delta$). For 
\begin{eqnarray} 
|\Omega|^2 \gg \gamma_0 \gamma_a, \;  \; \; \Delta \gg \gamma_{a},
 \Omega 
 \label{cond}
 \end{eqnarray}  
ideal correlations of Stokes and anti-Stokes fields are obtained.

For non-zero Fourier components the noise reduction deteriorates.
It is clear that the bandwidth of
squeezing is always on the order of the spectral width 
of the generated field, which becomes small near the oscillation threshold. 
For sufficiently
small $\gamma_0,\omega$ and $|M|$ we can approximate  $S(\omega)$ as:
\begin{eqnarray}  
S_{\rm opt}(\omega) \approx  
\frac{(|M|^2/2+ \sqrt{1+\omega^2/\delta \omega_0^2}-1)^2}{|M|^2 + 
\omega^2/\delta \omega_0^2} \nonumber \\  
+ {\pi \over 4} \left(2 
\frac{ \gamma_{0} \Delta}{ |\Omega|^2} 
+ 
\frac{ \gamma_a}{\Delta} 
\right ) Z(\omega),
\end{eqnarray}
where $\delta \omega_0 = \Omega^2/\Delta$ and we have set $\theta$
to the optimum value for each Fourier 
frequency $\theta =\pi/4 - 
\kappa L \omega/(2 \Omega^2)$. $Z(\omega) \sim 1$ is 
some function, which is  on the order of 
unity for arbitrary $\omega$. Its exact form is of no importance 
here. It follows 
directly from the 
above equation that squeezing is present 
for $\omega < \Omega^2/\Delta$, whereas maximum correlations, given by 
Eq. (\ref{ideal}),
occur within the bandwidth on the order of 
$\gamma_0 (\Omega^2 /\gamma_a \gamma_0)^{1/4}$.
It is worth noting that in the present system all relevant spectral widths are 
determined by the atomic dispersion \cite{harris98}. 

It is important to emphasize that the strong coupling regime
corresponding to the conditions of ideal photon correlations or squeezing
(\ref{cond})
can easily be realized even for very low driving-field 
intensities, since the ground-state relaxation rate 
$\gamma_{0}$ can be very small. For
example, in the experiments involving  hyperfine sublevels 
of the ground state of alkali  vapors such as \cite{zibrov98}, this
rate could be made as small as $1-100$ Hz. Even if a detuning 
$\Delta$ on the order of few tens of MHz is chosen, the Rabi frequencies
corresponding to the value $\Omega^2 = \gamma_{0}|\Delta |$
may well be close to the tens of kHz level. Under
experimental conditions where the  driving beams  are in single spatial 
modes and diffraction limited, 
$\lambda\sim 1\mu{\rm m}$, and where the 
optical pulse length is on the order of coherence life time
 ($\sim 1/\gamma_{0}$), the required Rabi frequencies 
correspond to only a few driving photons.
In this limit efficient parametric interactions and mixing involving 
only few interacting light quanta may take place. This opens up a rather 
unique regime of nonlinear optics which allows, at least in principle,  
for single photon quantum control 
\cite{harris98,quantum_control}, and for ``inelastic collisions'' of single 
light quanta
yielding correlated photons at different frequencies or 
polarizations. Furthermore, in such a regime the 
quantum nature of the driving 
fields as well as finite-size
effects may become important \cite{wro}.

In conclusion, we have demonstrated that resonant nonlinear
interactions involving atomic coherence can be used for efficient 
generation of quantum-correlated electromagnetic fields 
with  100\% squeezing without the use of cavities. 
We have shown that under appropriate conditions the resonant wave mixing
process based on double-$\Lambda$ atomic media can be regarded as
ideal even for extremely low driving input powers. 
We expect these features of coherent atomic systems to be of interest in many
areas of optics, spectroscopy and quantum control.

The authors gratefully acknowledge
useful discussions with 
L.~Hollberg, S.~Harris, P.~Hemmer, V.~Sautenkov,  and 
A.~Zibrov and the support from
the Office of Naval Research, the National Science Foundation, 
the Welch Foundation,
the Texas Advanced Research and Technology Program and the Air Force
Research  Laboratories.



\begin{figure}

\epsfxsize=6cm
\centerline{\epsffile{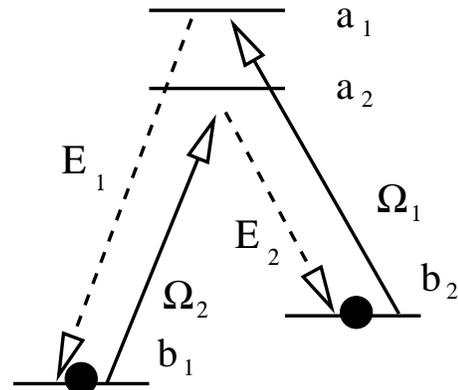}}

\caption{Atoms in double $\Lambda$ configuration interacting 
with two classical driving fields 
($\Omega_{1,2}$) and two quantum fields ($E_{1,2}$). All optical transitions
are assumed to be radiatively broadened.}
 \end{figure}



\begin{thebibliography}{99}
\bibitem{remark1}
Clearly at the threshold the linear approximation 
used in Eqs.(\ref{polar1}) breaks down and higher order corrections need 
 to be taken into account.
\bibitem{remark} Note that the $\delta$-function in space is due to 
vanishing corrleations between different atoms. The  
$\delta(\omega+\omega^\prime)$-term comes from  double Fourier transforming
$\delta(t-t^\prime)$ in $t$ and $t^\prime$.
\bibitem{wro} For these reasons direct 
 extrapolation of the present results into 
this domain is strictly speaking inconsistent.
\end{thebibliography}
\end{document}